\documentclass[sigconf]{acmart}

\usepackage[utf8]{inputenc}
\usepackage{booktabs,multirow}

\usepackage{dsfont}
\usepackage{bbm}
\usepackage{arydshln}
\usepackage{graphicx}   
\usepackage{subcaption} 

\hypersetup{
    colorlinks=true,
    linkcolor=blue,
    filecolor=magenta,      
    urlcolor=cyan,
    }

\usepackage[normalem]{ulem}
\usepackage{xcolor}
\usepackage[table]{xcolor}
\usepackage{color}
\usepackage{rotating}
\usepackage{amsmath}
\usepackage{array}
\usepackage{enumitem}
\usepackage[linesnumbered,algoruled,boxed,lined]{algorithm2e}
\usepackage{balance}
\usepackage[short]{optidef}
\usepackage{graphicx}
\usepackage{tcolorbox}
\usepackage{hyperref}

\newcolumntype{P}[1]{>{\centering\arraybackslash}p{#1}}
\newcolumntype{R}[1]{>{\RaggedLeft\arraybackslash}p{#1}}

\newcommand{\ie}{{\it i.e.}}
\newcommand{\eg}{{\it e.g.}}

\newcommand{\ul}{\underline}{}

\AtBeginDocument{%
  }

\setcopyright{acmcopyright}

\copyrightyear{2026}
\acmYear{2026}
\setcopyright{cc}
\setcctype{by-nc-nd}
\acmConference[SIGIR '26] {Proceedings of the 49th International ACM SIGIR Conference on Research and Development in Information Retrieval}{July 20--24, 2026}{Melbourne, VIC, Australia.}
\acmBooktitle{Proceedings of the 49th International ACM SIGIR Conference on Research and Development in Information Retrieval (SIGIR '26), July 20--24, 2026, Melbourne, VIC, Australia}
\acmISBN{979-8-4007-2599-9/2026/07}
\acmDOI{10.1145/3805712.3809911}

\title{ACE: Anisotropy-Controllable Embedding for LLM-enhanced Sequential Recommendation}



\author{Dongcheol Lee}\authornote{Equal contribution}
\affiliation{
  \institution{Sungkyunkwan University}
  \city{Suwon}
  \country{Republic of Korea}
  }
\email{ka05183@skku.edu}

\author{Hye-young Kim}\authornotemark[1]
\affiliation{
  \institution{Sungkyunkwan University}
  \city{Suwon}
  \country{Republic of Korea}
  }
\email{khyaa3966@skku.edu}

\author{Jongwuk Lee}\authornote{Corresponding author}
\affiliation{
  \institution{Sungkyunkwan University}
  \city{Suwon}
  \country{Republic of Korea}
  }
\email{jongwuklee@skku.edu}
\begin{document}

\begin{abstract}
Recent advances in the \emph{LLM-as-Extractor} paradigm leverage large language models (LLMs) to transfer semantically rich item embeddings into sequential recommendation (SR) backbones. However, LLM-generated embeddings often suffer from strong \emph{anisotropy}. Most vectors are concentrated in similar directions, resulting in a geometric imbalance that makes it difficult to adapt to collaborative signals during fine-tuning. To address this challenge, we propose \emph{Anisotropy-Controllable Embedding (\textbf{ACE})}, which explicitly controls the anisotropy of LLM-generated embeddings. Specifically, ACE utilizes a \emph{linear autoencoder (LAE)} to reshape the embedding distribution while preserving its semantic structure. In this process, the $L_2$-regularization term mitigates the anisotropy by controlling the dispersion of embedding dimensions, while the reconstruction loss maintains semantic relationships among items. That is, ACE balances geometric uniformity and semantic embedding preservation for more stable learning. Extensive experiments demonstrate that ACE consistently outperforms existing LLM-enhanced SR models, yielding improvements of up to 12.4\% and 11.8\% in Recall@20 and NDCG@20, respectively. 
\end{abstract}

\maketitle



\begin{figure}\small
    \centering
    \begin{subfigure}[t]{0.32\columnwidth}
        \centering
        \includegraphics[width=\linewidth]{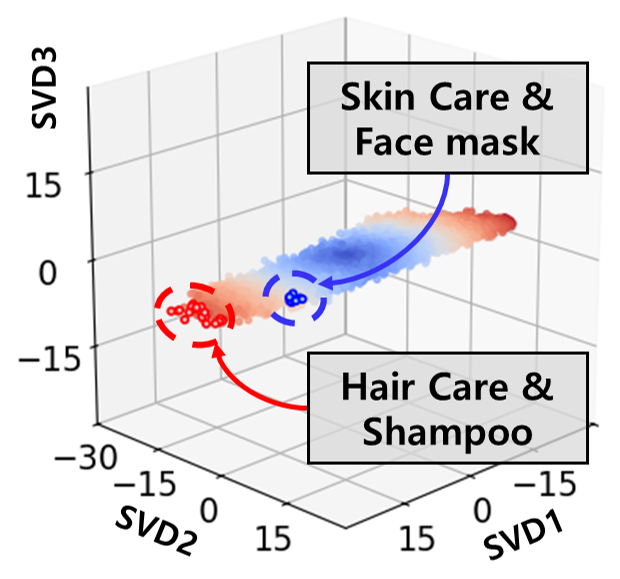}
        \caption{PCA}
        \label{fig:motivation_pca}
    \end{subfigure}
    \hfill
    \begin{subfigure}[t]{0.32\columnwidth}
        \centering
        \includegraphics[width=\columnwidth]{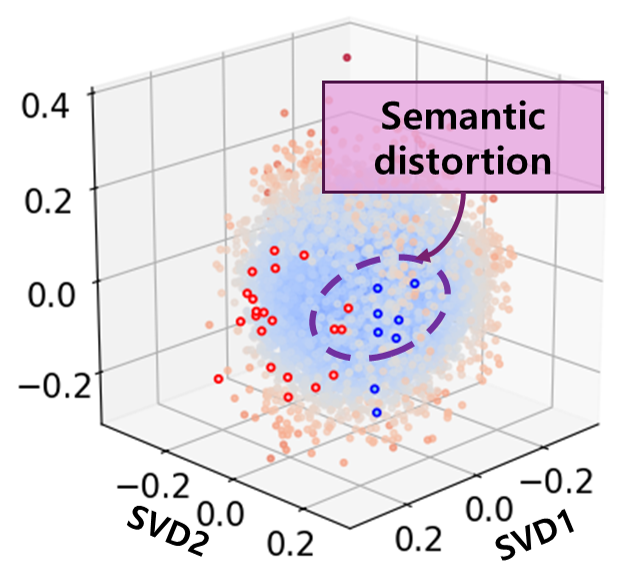}
        \caption{Whitened PCA}
        \label{fig:motivation_lae0}
    \end{subfigure}
    \hfill
    \begin{subfigure}[t]{0.32\columnwidth}
        \centering
        \includegraphics[width=\linewidth]{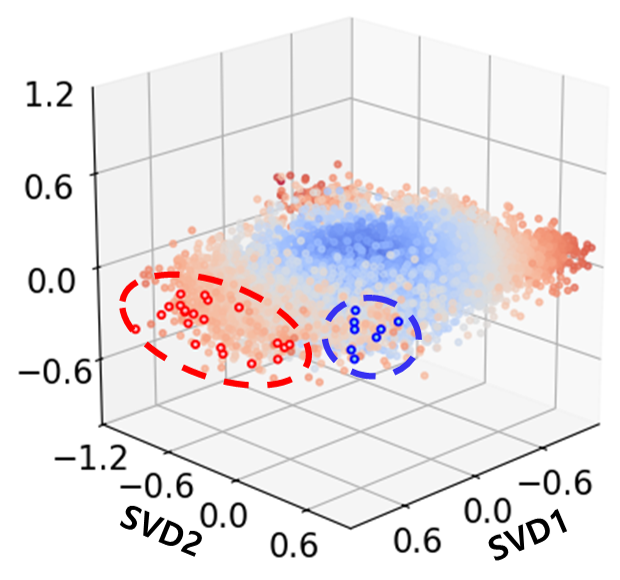}
        \caption{ACE (Ours)}
        \label{fig:motivation_lae50}
    \end{subfigure}
    \caption{
    Visualization of three embedding spaces on the Beauty dataset: PCA, whitened PCA, and ACE. Each point is projected onto the top three singular vectors (SVD1–3), where SVD1 and SVD3 correspond to semantically rich and sparse directions, respectively.}
    \vspace{-4mm}
    \label{fig:ACE_motivation}
\end{figure}

\section{Introduction}\label{sec:introduction}
Sequential recommendation (SR)~\cite{KangM18SASRec, HidasiKBT15GRU4Rec, SunLWPLOJ19BERT4Rec} focuses on predicting a user’s next preferred item by modeling their interaction history. Recently, Large language models (LLMs) have been integrated into sequential recommender (SR) models to enrich user and item representations with rich semantic knowledge. Notably, the high inference cost of LLMs has motivated the \textit{LLM-as-Extractor}~\cite{LLM2X, WWZ00025LLMEmb, AlphaRec, Hu0LC0025Alphafuse} paradigm, which substitutes randomly initialized item embeddings of conventional SR models with LLM-generated representations for downstream fine-tuning.

As an early study, LLM2X~\cite{LLM2X} transfers LLM-generated embeddings into SR models by applying a simple principal component analysis (PCA) followed by fine-tuning. However, this strategy provides limited control over how LLM semantic representations adapt to collaborative signals. Recent studies~\cite{HuXZFWHLTZ24SAID, WWZ00025LLMEmb, AlphaRec} have thus focused on improving the adaptation of LLM embeddings to SR backbones.
AlphaRec~\cite{AlphaRec}, LLMEmb~\cite{WWZ00025LLMEmb}, and WhitenRec~\cite{Zhang0Z024WhitenRec} employ MLP projection layers to align the LLM semantic space with collaborative signals. More recently, AlphaFuse~\cite{Hu0LC0025Alphafuse} freezes the semantically meaningful subspace while reinitializing the remaining dimensions to encourage collaborative learning. While these methods enhance adaptation to SR backbones, they largely overlook a more fundamental factor: the importance of properly initializing pretrained embeddings before fine-tuning.

In this paper, we investigate the geometry of LLM-generated embeddings with respect to \emph{anisotropy}~\cite{GaoHTQWL19anisotropy}. This property fundamentally influences how effectively LLM semantics adapt to collaborative signals during fine-tuning. Prior studies~\cite{Zhang0Z024WhitenRec} have shown that Sentence-BERT embeddings~\cite{ReimersG19sentence-bert} suffer from strong anisotropy, collapsing into a few dominant components. In Figure~\ref{fig:motivation_pca}, we observe that LLM-generated embeddings exhibit the same phenomenon. Most vectors are concentrated within a narrow cone, causing them to cluster tightly and lose representational diversity. To alleviate this, WhitenRec~\cite{Zhang0Z024WhitenRec} applies the whitening transformation that enforces perfect isotropy. However, this rigid operation collapses the eigenvalue hierarchy that reflects semantic importance, leading to substantial semantic distortion (Figure~\ref{fig:motivation_lae0}). These observations highlight the need for a more flexible and semantically faithful approach to addressing anisotropy in LLM-generated embeddings.

To this end, we introduce \textit{\textbf{A}nisotropy-\textbf{C}ontrollable \textbf{E}mbedding (\textbf{ACE})}, which provides continuous and semantically consistent control over the anisotropy of LLM-generated embeddings. ACE is built upon a \emph{linear autoencoder (LAE)}, which regulates the intrinsic spectral behavior of the embedding space. Through the spectral characterization of the LAE solution, we observe that its two objective components play distinct geometric roles: (i) the \textit{reconstruction objective} that preserves the semantic directions encoded in the original LLM embeddings, and (ii) the \textit{$\lambda$-weighted regularization term}, which directly controls the magnitude of singular values, suppressing overly dominant semantic directions and modulating anisotropy in a continuous manner. Together, ACE yields a well-conditioned embedding space that mitigates geometric imbalance while maintaining the underlying semantic hierarchy for stable learning.
As shown in Figure~\ref{fig:motivation_lae50}, it empirically reduces directional over-concentration without causing semantic distortion. Our extensive experiments further demonstrate that ACE consistently improves LLM-enhanced SR models, achieving performance gains of up to 12.4\% and 11.8\% in Recall@20 and NDCG@20, respectively.
\section{Preliminaries}\label{sec:preliminary}
\textbf{Problem Definition}.
Let $\mathcal{U}=\{u_1, \dots, u_m\}$ and $\mathcal{I}=\{i_1, \dots, i_n\}$ denote the sets of $m$ users and $n$ items. For an arbitrary user $u\in\mathcal{U}$, the sequence $S_u=[i_1, i_2, \text{...}, i_{|S_u|}]$ represents the items that the user has interacted with in chronological order. Given $S_u$, the goal of SR is to predict the next item $i_{|S_u|+1}$ that the user is likely to prefer. 

\vspace{1mm}
\noindent
\textbf{LLM-as-Extractor Paradigm}.
Recent advances in LLMs enable the usage of semantically rich item embeddings for SR. Given an item and its textual attributes, an LLM encoder produces an embedding $e \in \mathbb{R}^d$ for each item, and stacking these embeddings yields the item embedding matrix $\mathbf{E}= [e_1, e_2, …, e_n]^\top \in \mathbb{R}^{n \times d}$, which forms the LLM-generated representation space~\cite{Hu0LC0025Alphafuse}. These embeddings are typically high-dimensional (\eg, $d = 3072$) and therefore need to be projected into a lower-dimensional space (\eg, $k = 128$) using PCA~\cite{LLM2X} or MLP projection~\cite{AlphaRec, WWZ00025LLMEmb} for compatibility with SR models.

\begin{figure}\small
    \centering
    
    \includegraphics[width=0.85\linewidth]{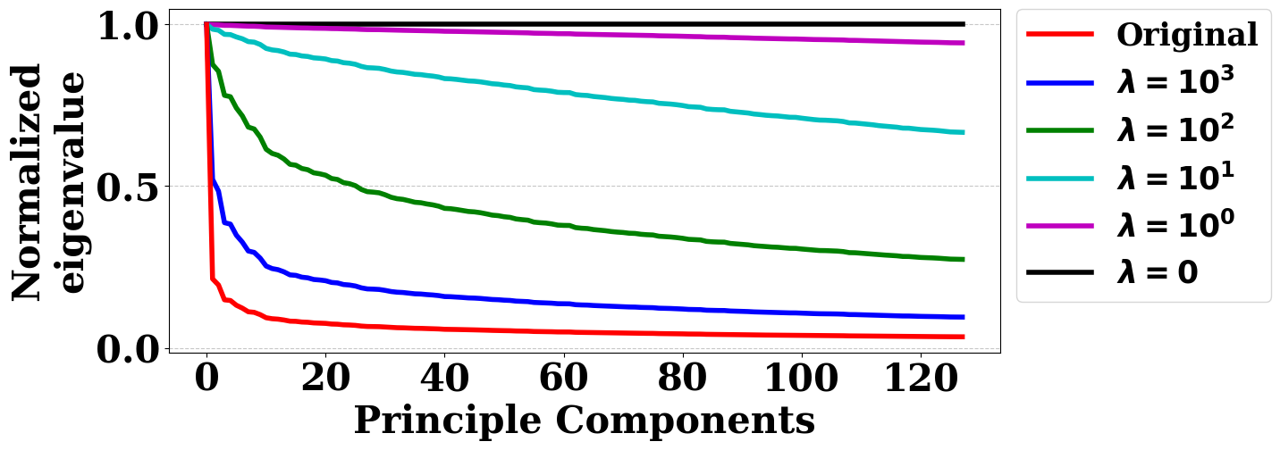}
    \vspace{-1mm}
    \caption{
        Normalized eigenvalue spectrum of ACE under different $\lambda$ on the Beauty dataset. Smaller $\lambda$ flattens the eigenvalue spectrum toward isotropy.
    }
    \vspace{-3mm}
    \label{fig:eigenvalue_fig}
\end{figure}

\vspace{1mm}
\noindent
\textbf{Embedding Anisotropy Problem}.
Despite the rich semantics, LLM-generated embeddings are known to be highly anisotropic~\cite{GaoHTQWL19anisotropy}, meaning that most embedding vectors lie in a narrow region of the space and align along a few dominant directions. Given the covariance matrix $\text{Cov}(\mathbf{E}) = \frac{1}{n}(\mathbf{E} - \mu)^\top (\mathbf{E} - \mu)$, where $\mu \in \mathbb{R}^{d}$ is the mean embedding vector, the distribution of its eigenvalues provides a direct measure of the anisotropy of embedding space. As shown in the original eigenvalue spectrum of Figure~\ref{fig:eigenvalue_fig} (\ie, red line), LLM embeddings reveal extreme variance imbalance. Such collapsed distributions distort embedding similarity and reduce semantic separability, which can harm downstream task performance~\cite{Zhang0Z024WhitenRec}.

\vspace{1mm}
\noindent
\textbf{Whitening-based Anisotropy Mitigation.}
To mitigate the embedding anisotropy problem, WhitenRec~\cite{Zhang0Z024WhitenRec} adopts the \textit{whitening transformation}~\cite{2000whitening}. Although whitening is commonly introduced through the eigendecomposition of a covariance matrix, it can be equivalently understood through the singular value decomposition (SVD), offering a clearer interpretation for LLM embeddings. 
Given the mean-centered embedding matrix $\tilde{\mathbf{E}}$, it can be decomposed as:
\begin{equation}
    \tilde{\mathbf{E}}=\tilde{\mathbf{U}} \tilde{\mathbf{S}} \tilde{\mathbf{V}}^\top,
\end{equation}
where $\tilde{\mathbf{V}}$ contains the principal components and $\tilde{\mathbf{S}}$ is the diagonal matrix of singular values. A strong anisotropy manifests as a highly skewed singular value spectrum, indicating that most variance is confined to a few directions.
From the SVD perspective, whitening rescales each principal component by the inverse singular value~\cite{2000whitening}:
\begin{equation}
    \mathbf{E}_\text{whiten} = \tilde{\mathbf{E}} \tilde{\mathbf{V}} {\tilde{\mathbf{S}}^{-1}} = \tilde{\mathbf{U}},
\end{equation}
resulting in whitened features with an identity covariance matrix, \ie, $\text{Cov}(\mathbf{E}_{\text{whiten}}) = \mathbf{I}$. This transformation equalizes variance across all directions, thereby enforcing perfect isotropy. However, this isotropic property of whitening collapses the representational spectrum. WhitenRec+~\cite{Zhang0Z024WhitenRec} applies group-wise whitening to preserve some semantic coherence, but its discrete formulation still lacks continuous control and preserves the internal semantic hierarchy.

\section{Proposed Method: ACE}\label{sec:model}
In this section, we present \textit{Anisotropy-Controllable Embedding (\textbf{ACE})}, which mitigates the geometric imbalance of LLM-generated embeddings. ACE employs a \textit{linear autoencoder (LAE)}, whose spectral decomposition enables continuous control of anisotropy while preserving the semantic hierarchy of LLM embeddings.

\vspace{1mm}
\noindent
\textbf{LAE-based Embedding Reshaping}.
ACE applies a Linear Autoencoder~\cite{Steck19EASE} (LAE) to reshape LLM-generated item embeddings prior to training the SR backbones. We directly use the LLM embedding matrix $\mathbf{E} \in \mathbb{R}^{n\times d}$ as the input and define an LAE objective that adjusts its spectral structure. To this end, we introduce an item–item similarity matrix $\mathbf{B}_{\text{ACE}} \in \mathbb{R}^{n\times n}$, and the ACE objective:
\begin{equation}
\mathcal{L}(\mathbf{B}_\text{ACE}) 
= \min_{\mathbf{B}_\text{ACE}} 
\| \mathbf{E}^\top - \mathbf{E}^\top\mathbf{B}_\text{ACE} \|_F^2 
+ \lambda \| \mathbf{B}_\text{ACE} \|_F^2,
\label{eq:lae_llm}
\end{equation}
where the first term reconstructs the original embeddings, and the second term regularizes the $\mathbf{B}_\text{ACE}$ matrix to avoid trivial solutions~\cite{Steck19EASE, MoonKL23RDLAE}. Here, $\lambda$ is a hyperparameter to control the strength of the second term (\ie, regularization). This objective yields a closed-form solution:
\begin{equation}
    \hat{\mathbf{B}}_{\text{ACE}} = (\mathbf{E}\mathbf{E}^{\top} + \lambda \textbf{I})^{-1}\mathbf{E}\mathbf{E}^{\top}.
    \label{eq:lae_closed}
\end{equation}

\begin{table*}[]
\small
\centering
\caption{Performance comparison with LLM-as-Extractor baselines across four benchmark datasets using SASRec backbone. The best results are shown in bold, and the second-best are \ul{underlined}. `*' denotes a statistically significant improvement over all baselines ($p < 0.05$, one-tailed t-test).}
\renewcommand{\arraystretch}{0.9}
\setlength{\tabcolsep}{1.6pt} 
\label{tab:overall_extended}
\vspace{-3mm}
\begin{tabular}{c|cccc|cccc|cccc|cccc}
\toprule
                        & \multicolumn{4}{c|}{\textbf{Beauty}}       & \multicolumn{4}{c|}{\textbf{Toys}}         & \multicolumn{4}{c|}{\textbf{Yelp}}         & \multicolumn{4}{c}{\textbf{ML-20M}}        \\
\multirow{-2}{*}{Model} & R@10   & N@10   & R@20   & N@20   & R@10   & N@10   & R@20   & N@20   & R@10   & N@10   & R@20   & N@20   & R@10   & N@10   & R@20   & N@20   \\ \midrule
SASRec                  & 0.0498 & 0.0255 & 0.0705 & 0.0307 & 0.0465 & 0.0253 & 0.0596 & 0.0285 & 0.0368 & 0.0256 & 0.0489 & 0.0286 & 0.1893 & 0.1018 & 0.2643 & 0.1207 \\ \midrule
LLM2X                   & 0.0827 & 0.0403 & 0.1181 & 0.0492 & 0.0881 & 0.0437 & 0.1241 & 0.0527 & 0.0590 & 0.0356 & 0.0851 & 0.0422 & 0.1916 & 0.1053 & 0.2660 & 0.1240 \\
WhitenRec+              & 0.0781 & 0.0388 & 0.1184 & 0.0489 & 0.0869 & 0.0423 & 0.1268 & 0.0524 & 0.0500 & 0.0260 & 0.0799 & 0.0335 & 0.1973 & 0.1095 & 0.2765 & 0.1294 \\
LLMEmb                  & 0.0782 & 0.0372 & 0.1144 & 0.0463 & 0.0846 & 0.0396 & 0.1218 & 0.0490 & 0.0539 & 0.0305 & 0.0810 & 0.0373 & 0.1973 & 0.1087 & 0.2767 & 0.1287 \\
AlphaRec                & 0.0773 & 0.0384 & 0.1167 & 0.0482 & 0.0873 & 0.0428 & 0.1292 & 0.0534 & 0.0535 & 0.0281 & 0.0832 & 0.0355 & 0.1992 & \ul{0.1096} & \ul{0.2810} & \ul{0.1302} \\
AlphaFuse               & \ul{0.0864} & \ul{0.0422} & \ul{0.1263} & \ul{0.0523} & \ul{0.0931} & \ul{0.0442} & \ul{0.1317} & \ul{0.0540} & \ul{0.0604} & \ul{0.0359} & \ul{0.0879} & \ul{0.0428} & \ul{0.2003} & 0.1092 & 0.2798 & 0.1292 \\ \midrule
\rowcolor{blue!18!gray!20} 
\textbf{ACE (Ours)}              & \textbf{0.0886*} & \textbf{0.0433*} & \textbf{0.1288*} & \textbf{0.0534*} & \textbf{0.0977*} & \textbf{0.0468*} & \textbf{0.1342*} & \textbf{0.0560*} & \textbf{0.0610} & \textbf{0.0384*} & \textbf{0.0892*} & \textbf{0.0455*} & \textbf{0.2032*} & \textbf{0.1110} & \textbf{0.2825*} & \textbf{0.1310*} \\ \bottomrule
\end{tabular}
\end{table*}
ACE aims to directly construct a geometrically adjusted embedding matrix $\mathbf{E}_\text{ACE}$ that mitigates embedding anisotropy. Since anisotropy arises from an imbalanced singular value spectrum, we start from the singular value decomposition of the original embedding matrix $\mathbf{E}=\mathbf{U}\mathbf{S}\mathbf{V}^{\top}$. We define the ACE-adjusted embedding $\mathbf{E}_\text{ACE}$ by applying a shrinkage function to each singular value: 
\begin{equation}
    \mathbf{E}_\text{ACE} = \mathbf{U} \, g_\lambda(\mathbf{S}),
    \label{eq:lae_embed_s}
\end{equation}
where $\mathbf{U}$ contains the principal directions of the item-item geometry induced by $\mathbf{E}\mathbf{E}^{\top}$ and the shrinkage function is defined as:
\begin{equation}
g_\lambda(\mathbf{S}) = \sqrt{\frac{\mathbf{S}^2}{\mathbf{S}^2 + \lambda \mathbf{I}}}=\mathrm{diag}\!\left(
    \sqrt{\frac{\sigma_1^2}{\sigma_1^2 + \lambda}}, \ldots,
    \sqrt{\frac{\sigma_n^2}{\sigma_n^2 + \lambda}}
    \right).
\label{eq:singular_value}
\end{equation}

Note that $\mathbf{V}$ is omitted since ACE focuses on reshaping the item–item similarity geometry rather than reconstructing the original feature space. To mitigate anisotropy, ACE attenuates overly dominant singular directions while preserving relative semantic structure, achieved via magnitude-dependent spectral shrinkage.

The corresponding linear operator induced by $\mathbf{E}_\text{ACE}$ is then $\hat{\mathbf{B}}_{\text{ACE}}=\mathbf{E}_\text{ACE}\mathbf{E}_\text{ACE}^\top$,  which yields the following spectral form:
\begin{equation}
    \hat{\mathbf{B}}_{\text{ACE}} 
    = \mathbf{U}\, \frac{\mathbf{S}^2}{\mathbf{S}^2+\lambda \mathbf{I}}\, \mathbf{U}^{\top} 
    =\mathbf{U}\,\mathrm{diag}\!\left(
    \frac{\sigma_1^2}{\sigma_1^2 + \lambda}, \ldots,
    \frac{\sigma_n^2}{\sigma_n^2 + \lambda}
    \right)\,\mathbf{U}^{\top},
    \label{eq:lae_svd2}
\end{equation}
which coincides with the spectral solution derived in prior work~\cite{MoonKL23RDLAE}.

\vspace{1mm}
\noindent
\textbf{Interpretation of ACE Objective}.
Based on this SVD characterization, we now discuss the roles of the two components in the ACE objective and how each contributes to controlling anisotropy.

(i) \textit{Reconstruction loss} (\ie, $\|\mathbf{E}^\top - \mathbf{E}^\top\mathbf{B}\|_F^2$) is responsible for preserving the semantic directions encoded by the original LLM embeddings. When $\lambda=0$, the second term (\ie, $L_2$-regularization) in Eq.~\eqref{eq:lae_llm} disappears, and ACE reduces to optimizing only the reconstruction loss. Substituting $\lambda=0$ into the shrinkage function in Eq.~\eqref{eq:singular_value} yields $g_{0}(\mathbf{\sigma_i})=1$ for all $i$, meaning that all singular components are retained with equal weight.
Consequently, the reconstructed embedding matrix $\mathbf{E}_\text{ACE}=\mathbf{U}$ is expressed entirely in the left singular basis $\mathbf{U}$, which represents the semantic directions of the original LLM embedding space. Thus, the reconstruction loss of ACE preserves semantic directions exactly, an effect analogous to whitening, \ie, every principal component is normalized to equal variance 1.

(ii) \textit{$L_2$-regularization term} (\ie, $\lambda\|\mathbf{B}\|_F^2$) governs how strongly each principal component is weighted by regulating the shrinkage function $g_{\lambda}(\mathbf{S})$ in Eq.~\eqref{eq:singular_value}. Figure ~\ref{fig:eigenvalue_fig} illustrates how the shrinkage function progressively smooths the singular value spectrum as $\lambda$ increases, demonstrating the continuous anisotropy control enabled by ACE. Unlike the $\lambda=0$ case, where all directions are treated equally and leading to an isotropic embedding space, increasing $\lambda$ gradually downweights dominant singular values and smooths the spectral imbalance. When $\lambda$ is very large, $g_{\lambda}(\mathbf{S})\approx \mathbf{S}/\sqrt{\lambda}$, which follows the original eigenvalue distribution of the LLM embeddings.
This makes the basis vectors in $\mathbf{U}$ weighted according to their original variance, thus maintaining an anisotropic representation. In this way, the regularization term acts as a \emph{generalized form of whitening}, modulating the spectral balance of the embedding space without collapsing its semantic hierarchy.

\vspace{1mm}
\noindent
\textbf{Applying ACE to LLM-as-Extractor}.
After the embedding reshaping, ACE integrates seamlessly into the LLM-as-Extractor paradigm by producing a geometry-adjusted, $k$-dimensional embedding matrix used to initialize the SR backbones. 

Given the spectral formulation in Eq.~\eqref{eq:lae_embed_s}, ACE constructs the reduced embedding matrix by retaining the top-$k$ singular directions:
\begin{equation}
    \mathbf{E}_\text{ACE}^{(k)} = \mathbf{U}^{(k)}\, g_\lambda(\mathbf{S}^{(k)}) \cdot \gamma,
\end{equation}
where $\mathbf{U^{(k)}}$ and $\mathbf{S^{(k)}}$ denote the top-$k$ singular vectors and values, respectively. Since $g_\lambda(\cdot)$ compresses the embedding magnitudes, ACE introduces a scaling factor $\gamma$ to recover the embedding scale.

The resulting normalized embeddings serve as the initialization of the item embedding table in any SR backbone, \eg, SASRec~\cite{KangM18SASRec}, GRU4Rec~\cite{HidasiKBT15GRU4Rec}, or BERT4Rec~\cite{SunLWPLOJ19BERT4Rec}.
The SR model is then trained as usual, benefiting from an embedding space that is both semantically coherent and geometrically balanced.


\section{Experimental Setup}
\textbf{Datasets}.
We evaluate ACE on four benchmark datasets: two Amazon Review 2014 subcategories (Beauty and Toys) ~\cite{sigir/McAuleyTSH15Amazonreview}, the Yelp 2018 dataset\footnote{\url{https://business.yelp.com/data/resources/open-dataset/}}, and the ML-20M dataset\footnote{\url{https://grouplens.org/datasets/movielens/20m/}}. Following existing works~\cite{WWZ00025LLMEmb, Hu0LC0025Alphafuse}, we filter out users and items with fewer than five interactions. For LLM-based item initialization, we encode each item’s textual metadata (\eg, title, category, brand) using \textit{text-embedding-3-large} from OpenAI\footnote{\url{https://platform.openai.com/docs/guides/embeddings}}. In addition, we experiment with multiple LLM encoders for item initialization, including \textit{F2LLM-4B}~\cite{2025F2LLM}, \textit{Qwen3-Embedding-8B}~\cite{qwen3embedding}, and \textit{KaLM-Embedding-Gemma3-12B-2511}~\cite{hu2025kalmembedding, zhao2025kalmembeddingv2}.

\vspace{1mm}
\noindent
\textbf{Evaluation Protocols}.
Following~\cite{Hu0LC0025Alphafuse}, we adopt the leave-one-out protocol to construct training, validation, and test sequences for each user. For evaluation, we rank all items in a full-sort manner without additional candidate filtering, and report Recall@$K$ (R@$K$) and NDCG@$K$ (N@$K$), where $K \in \{10,20\}$.

\vspace{1mm}
\noindent
\textbf{Baseline Models}.
We compare ACE with five competitive LLM-as-Extractor approaches. PCA-based method (LLM2X~\cite{LLM2X}), whitening-based methods (WhitenRec+~\cite{Zhang0Z024WhitenRec}), and semantic adaptation approaches (LLMEmb~\cite{WWZ00025LLMEmb}, AlphaRec~\cite{AlphaRec}, and AlphaFuse~\cite{Hu0LC0025Alphafuse}). For integrating LLM-as-Extractor approaches into SR backbones, we employ three widely used SASRec~\cite{KangM18SASRec}, GRU4Rec~\cite{HidasiKBT15GRU4Rec}, and BERT4Rec~\cite{SunLWPLOJ19BERT4Rec}.

\vspace{1mm}
\noindent
\textbf{Implementation Details}.
All models are implemented using the open-source RecBole framework~\cite{ZhaoMHLCPLLWTMF21RecBole}. We set the item embedding dimension to 128, the maximum sequence length to 50, and use a batch size of 256. We adopt early stopping with a patience of 10 epochs, based on NDCG@10 on the validation set. All models are trained with Adam optimizer and the learning rate is tuned on a logarithmic scale from $10^{-3}$ to $10^{-5}$. For ACE, the $L_2$-regularization coefficient $\lambda$ is selected from $\{0,1,5, \cdots,  5\times10^3\}$, and the scaling parameter $\gamma$ is used to rescale the embedding standard deviation to $\{0.1, 0.5, 1\}$. All results are averaged over four runs. The source code is available at \href{https://github.com/DCheol/ACE}{https://github.com/DCheol/ACE}.

\begin{table}[]
\small
\centering
\caption{Performance comparison against LLM-as-Extractor baselines across two SR backbones on the Beauty and Yelp datasets. The best performance is highlighted in bold, and the second-best is \ul{underlined}.}
\vspace{-1mm}
\renewcommand{\arraystretch}{1}
\label{table:another_model}

\begin{tabular}{c|c|cc|cc}
\toprule
 &
   &
  \multicolumn{2}{c|}{\textbf{GRU4Rec}} &
  \multicolumn{2}{c}{\textbf{BERT4Rec}} \\
\multirow{-2}{*}{Dataset} &
  \multirow{-2}{*}{Model} &
  R@20 &
  N@20 &
  R@20 &
  N@20 \\ \midrule
 &
  Original &
  0.0537 &
  0.0244 &
  0.0384 &
  0.0157 \\ \cmidrule{2-6}
 &
  LLM2X &
  0.0758 &
  0.0318 &
  0.0737 &
  0.0297 \\
 &
  WhitenRec+ &
  0.1022 &
  0.0428 &
  0.0839 &
  0.0335 \\
 &
  LLMEmb &
  0.0907 &
  0.0371 &
  0.0857 &
  0.0344 \\
 &
  AlphaRec &
  {\ul {0.1025}} &
  {\ul {0.0430}} &
  {\ul {0.0921}} &
  {\ul {0.0369}} \\
 &
  AlphaFuse &
  0.0992 &
  0.0420 &
  0.0841 &
  0.0333 \\ \cmidrule{2-6}
\multirow{-8}{*}{Beauty} &
  \cellcolor{blue!18!gray!20}\textbf{ACE (Ours)} &
  \cellcolor{blue!18!gray!20}\textbf{0.1107} &
  \cellcolor{blue!18!gray!20}\textbf{0.0466} &
  \cellcolor{blue!18!gray!20}\textbf{0.0926} &
  \cellcolor{blue!18!gray!20}\textbf{0.0372} \\ \midrule
 &
  Original &
  \multicolumn{1}{r}{0.0358} &
  \multicolumn{1}{r|}{0.0193} &
  \multicolumn{1}{r}{0.0355} &
  \multicolumn{1}{r}{0.0138} \\ \cmidrule{2-6}
 &
  LLM2X &
  \multicolumn{1}{r}{0.0454} &
  \multicolumn{1}{r|}{0.0191} &
  \multicolumn{1}{r}{0.0726} &
  \multicolumn{1}{r}{0.0293} \\
 &
  WhitenRec+ &
  \multicolumn{1}{r}{{\ul {0.0765}}} &
  \multicolumn{1}{r|}{0.0314} &
  \multicolumn{1}{r}{0.0713} &
  \multicolumn{1}{r}{0.0281} \\
 &
  LLMEmb &
  \multicolumn{1}{r}{0.0675} &
  \multicolumn{1}{r|}{0.0282} &
  \multicolumn{1}{r}{0.0728} &
  \multicolumn{1}{r}{0.0290} \\
 &
  AlphaRec &
  \multicolumn{1}{r}{0.0762} &
  \multicolumn{1}{r|}{0.0311} &
  \multicolumn{1}{r}{0.0756} &
  \multicolumn{1}{r}{0.0303} \\
 &
  AlphaFuse &
  \multicolumn{1}{r}{0.0762} &
  \multicolumn{1}{r|}{{\ul {0.0339}}} &
  \multicolumn{1}{r}{{\ul {0.0811}}} &
  \multicolumn{1}{r}{{\ul {0.0333}}} \\ \cmidrule{2-6}
\multirow{-8}{*}{Yelp} &
  \cellcolor{blue!18!gray!20}\textbf{ACE (Ours)} &
  \multicolumn{1}{r}{\cellcolor{blue!18!gray!20}\textbf{0.0860}} &
  \multicolumn{1}{r|}{\cellcolor{blue!18!gray!20}\textbf{0.0379}} &
  \multicolumn{1}{r}{\cellcolor{blue!18!gray!20}\textbf{0.0838}} &
  \multicolumn{1}{r}{\cellcolor{blue!18!gray!20}\textbf{0.0346}} \\\bottomrule
\end{tabular}
\vspace{-2mm}
\end{table}

\section{Experimental Results}\label{sec:results}
\textbf{Overall Performance}.
Table~\ref{tab:overall_extended} shows that ACE consistently outperforms all existing LLM-as-Extractor models across all benchmark datasets and backbone architectures. Notably, ACE achieves absolute improvements of up to 4.9\% in Recall@10 and 5.9\% in NDCG@10 over the best competing baseline. Notably, whitening-based method (\ie, WhitenRec+~\cite{Zhang0Z024WhitenRec}) sometimes underperforms even the simple PCA-based LLM2X~\cite{LLM2X}. This degradation aligns with our geometric analysis: aggressive whitening uniformly flattens the embedding spectrum, distorting the semantic hierarchy essential for item differentiation. In contrast, ACE provides a more balanced embedding geometry, achieving substantial gains by simultaneously reducing anisotropy and preserving semantic hierarchy.

\vspace{1mm}
\noindent
\textbf{Various Backbone Analysis}.
To verify that ACE generalizes beyond a specific SR architecture, we further evaluate it on two representative SR backbones, GRU4Rec~\cite{HidasiKBT15GRU4Rec} and BERT4Rec~\cite{SunLWPLOJ19BERT4Rec}. As shown in Table~\ref{table:another_model}, ACE consistently delivers performance gains across both backbones, improving Recall@20 by up to 12.4\% and NDCG@20 by up to 11.8\% over the strongest LLM-as-Extractor baseline. These improvements are observed across all datasets, indicating that ACE’s spectral reshaping mechanism remains effective regardless of sequence modeling architecture. These results highlight that ACE provides a robust and architecture-agnostic embedding reshaping.

\begin{figure}\small
    \centering
    \includegraphics[width=0.99\linewidth]{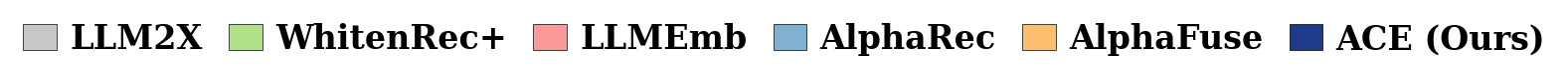}
    \includegraphics[width=0.75\linewidth]{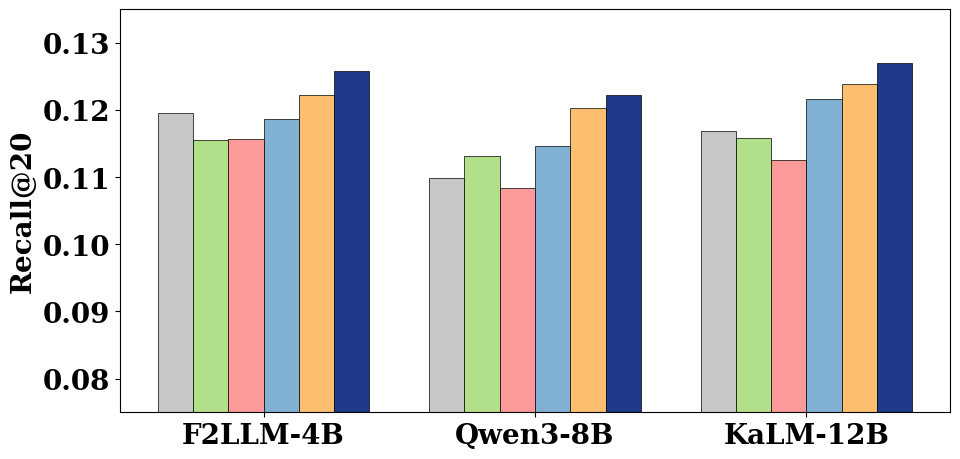}
    \vspace{-1mm}
    \caption{
        Performance comparison across three different LLM encoders: F2LLM-4B, Qwen3-Embedding-8B (Qwen3-8B), and KaLM-Embedding-Gemma3-12B-2511 (KaLM-12B), using SASRec backbone on the Beauty dataset.
    }
    \vspace{-1mm}
    \label{fig:llmsource}
\end{figure}

\begin{figure}\small
    \centering
    \begin{subfigure}[b]{0.49\columnwidth}
        \centering
        \includegraphics[width=\linewidth]{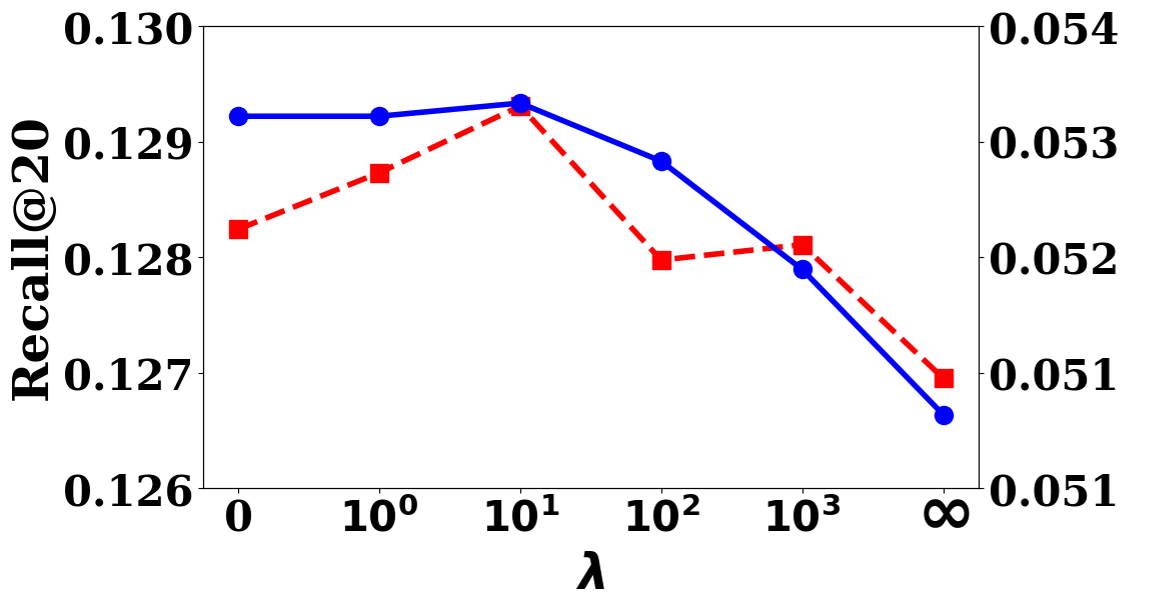}
        \caption{Beauty}
        \label{fig:hyper_lambda_beauty}
    \end{subfigure}
    \hfill
    \begin{subfigure}[b]{0.49\columnwidth}
        \centering
        \includegraphics[width=\linewidth]{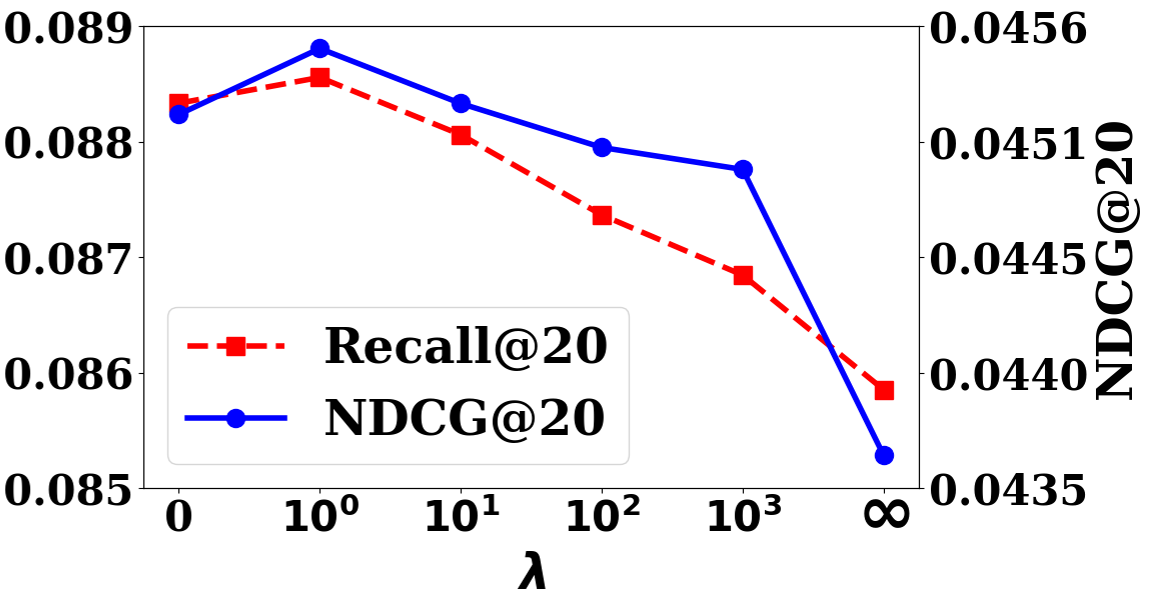}
        \caption{Yelp}
        \label{fig:hyper_lambda_yelp}
    \end{subfigure}
    \vspace{-1mm}
    \caption{
    Performance of ACE with varying $L_2$-regularization coefficient $\lambda$ using SASRec backbone on two datasets.
    }
    \vspace{-2mm}
    \label{fig:hyperparam_fig}
\end{figure}
\vspace{1mm}
\noindent
\textbf{Various LLM Encoders}.
Figure~\ref{fig:llmsource} reports the results on the Beauty dataset for LLM-as-Extractor baselines using SASRec backbone under different LLM encoders: \textit{F2LLM-4B}~\cite{2025F2LLM}, \textit{Qwen3-Embedding-8B}~\cite{qwen3embedding}, and \textit{KaLM-Embedding-Gemma3-12B-2511}~\cite{hu2025kalmembedding, zhao2025kalmembeddingv2}. Across all encoder settings, ACE consistently achieves the best performance on Recall@20, outperforming all LLM-as-Extractor baselines. Specifically, ACE improves Recall@20 by up to 2.94\% over the strongest baseline (\ie, AlphaFuse) under each LLM encoder. ACE maintains stable and consistent gains across all encoders, indicating that its performance gains are not tied to a specific LLM embedding.

\vspace{1mm}
\noindent
\textbf{Hyperparameter Analysis}.
Figure~\ref{fig:hyperparam_fig} illustrates how the performance of ACE with the SASRec backbone varies with the regularization coefficient $\lambda$. When $\lambda$ is too small, the shrinkage function over-flattens the spectrum, erasing semantic hierarchy and degrading accuracy. Conversely, very large $\lambda$ values provide insufficient shrinkage, leaving the original anisotropy largely unchanged and limiting representation quality. The best results consistently emerge at moderate $\lambda$ values, which balance anisotropy reduction with semantic hierarchy preservation. This demonstrates the effectiveness of ACE’s balanced anisotropy-control mechanism.

\section{Conclusion}\label{sec:conclusion}
In this work, we investigate the geometric limitations of LLM-generated embeddings, highlighting that their strong anisotropy hinders effective adaptation to collaborative signals. To address this issue, we introduced \textit{Anisotropy-Controllable Embedding (\textbf{ACE})}, which reshapes LLM embeddings through the spectral behavior of a \textit{linear autoencoder (LAE)}. By preserving semantic directions through reconstruction loss and modulating spectral variance via an $L_2$-regularization term, ACE provides continuous and controllable anisotropy mitigation. This yields geometrically balanced embeddings that maintain semantic integrity while effectively reducing anisotropy. Our experimental results confirm that ACE delivers consistent improvements, achieving gains of up to 12.4\% in Recall@20 and 11.8\% in NDCG@20, respectively.


\begin{acks}
    This work was partly supported by the Institute of Information \& communications Technology Planning \& evaluation (IITP) grant and the National Research Foundation of Korea (NRF) grant funded by the Korea government (MSIT) (No. RS-2022-II220680, RS-2025-00564083, RS-2019-II190421, RS-2024-00360227, each contributing 25\% to this research).
\end{acks}

\bibliographystyle{ACM-Reference-Format}
\balance
\bibliography{references}

\end{document}